\documentclass[aps,preprint,showpacs,preprintnumbers,amsmath,amssymb]{revtex4}
\usepackage{amsmath,mathrsfs,amsbsy,color,graphicx,bm,amsthm,amsfonts}
\usepackage{units}
\usepackage{bbm}
\usepackage{times}
\usepackage{dcolumn}
\usepackage{mathrsfs}% Align table columns on decimal point
\usepackage{amsmath,amssymb,epsfig}

\def\aap{Astron. Astrophys.}

\def\apjl{Astrophys. J. Lett.}

\def\apjs{Astrophys. J., Suppl. Ser.}

\begin{document}

\title{What are recent observations telling us in light of improved tests of distance duality relation?}
\author{Tonghua Liu$^{1,2}$, Shuo Cao$^{2,3}$ \footnote{Corresponding author: caoshuo@bnu.edu.cn}, Shuai Ma$^4$, Yuting Liu$^{2,3}$, Chenfa Zheng$^{2,3}$, Jieci Wang$^{5}$ \footnote{Corresponding author: jcwang@hunnu.edu.cn}}
\affiliation{1. School of Physics and Optoelectronic, Yangtze
University, Jingzhou 434023, China; \\ 2. Institute for Frontiers in Astronomy and Astrophysics, Beijing Normal University, Beijing 102206, China; \\
3. Department of Astronomy, Beijing Normal
University, Beijing 100875, China;\\
4. Beijing Academy, Kangyuan Road, Dongba, Beijing 100018, China;\\
5. Department of Physics, and Collaborative Innovation Center for Quantum Effects
and Applications, Hunan Normal University, Changsha, Hunan 410081, China.}

 \baselineskip=0.65 cm

\vspace*{0.2cm}
\begin{abstract}
As an exact result required by the Etherington reciprocity theorem, the cosmic distance duality relation (CDDR), $\eta(z)=D_L(z)(1+z)^{-2}/D_A(z)=1$ plays an essential part in modern cosmology. In this paper, we present a new method ($\eta(z_i)/\eta(z_j)$) to use the measurements of ultra-compact structure in radio quasars (QSO) and the latest observations of type Ia supernova (SN Ia) to test CDDR. By taking the observations directly from SN Ia and QSOs, one can completely eliminate the uncertainty caused by the calibration of the absolute magnitudes of standard candles ($M_B$) and the linear sizes of standard rulers ($l_m$). Benefit from the absence of nuisance parameters involved in other currently available methods, our analysis demonstrates no evidence for the deviation and redshift evolution of CDDR up to $z=2.3$. The combination of our methodology and the machine learning Artificial Neural Network (ANN) would produce $10^{-3}$ level constraints on the violation parameter at high redshifts. Our results indicate perfect agreement between observations and predictions, supporting the persisting claims that the Etherington reciprocity theorem could still be the best description of our universe.
\end{abstract}

\vspace*{0.5cm}
\maketitle

\section{Introduction}

The cosmic distance duality relation (CDDR), as a fundamental relation in modern cosmology, correlates the luminosity distance $D_L(z)$ with the angular diameter distance $D_A(z)$. More specifically, the CDDR indicates that $D_L(z)$ and $D_A(z)$ should satisfy the relation of $D_L(z)=D_A(z)(1+z)^2$ at the same redshift \cite{1933PMag...15..761E,2007GReGr..39.1055E}. However, the validity of the CDDR depends on three basic assumptions: i) the space-time is described by metric; ii) the light travels along the null geodesics between the observer and the source; iii) the photon number is conserved, i.e., the CDDR will be violated if the number of photons in the universe is not conserved. Therefore, the validity test of CDDR is, to some extent, an indirect verification of the accelerated expansion of the universe \citep{1998AJ....116.1009R,1999ApJ...517..565P,2004ApJ...607..665R,2003ApJ...598..102K}. In addition,  as a fundamental relationship in cosmology, the CDDR has been widely used in various fields of astronomy, such as the observations of large-scale distribution of galaxies and the near-uniformity of  the cosmic microwave background radiation (CMBR) temperature \cite{2020A&A...641A...6P}, the determination of gas mass density profile and temperature profile of galaxy clusters \cite{2011SCPMA..54.2260C,2016MNRAS.457..281C}, as well as the measurements of cosmic curvature with strong gravitational lensing systems \cite{2020MNRAS.496..708L,2017ApJ...834...75X,2019MNRAS.483.1104Q}.

On the other hand, the accumulation of precise astrophysical observations allow us to test the validity of CDDR at different redshifts. From the theoretical point of view, two types of cosmological distances are usually required in developing CDDR tests, i.e., angular diameter distance $D_A$ and luminosity distance $D_L$. For the observations of luminosity distances, many works turn to luminous sources with known (or standardizable) intrinsic luminosity in the universe like type-Ia supernova (SN Ia), while the angular diameter distances are inferred from baryon acoustic oscillations (BAO), Sunyaev-Zeldovich (SZ) effect of galaxy clusters \cite{2015PhRvD..92b3520W,2010ApJ...722L.233H,2011RAA....11.1199C}, gas mass fraction measurements in galaxy clusters \cite{2021JCAP...06..052B,2012ApJ...745...98M}, and strong gravitational lensing systems \cite{2018ApJ...866...31R,2019ApJ...885...70L,2019ApJ...887..163M,2022ChPhL..39k9801L}. However, it is worth noting that angular diameter distances derived from BAO is puzzled by the so-called fitting problem, which is a major challenge confronted by the standard BAO peak location with a fixed comoving ruler of about 105 h$^{-1}$ Mpc \cite{1987CQGra...4.1697E}. Therefore, such distance estimation is model-dependent to some extent, which inevitably brings systematic uncertainties and further affects the validity of testing CDDR with BAO. Meanwhile, galaxy clusters alone are not able to provide a competitive source of angular diameter distance at different redshifts, suffering from the large observational uncertainties arising from radio observations of the SZ effect of galaxy clusters together with X-ray emission. In addition, based on the observations of SN Ia, it was argued that the nuisance parameters characterizing SN Ia light-curves also introduce considerable uncertainties to the final results \cite{2016ApJ...833..240L,2017ApJ...838..160W}. Therefore,
in order to perform the validity of testing CDDR, one needs to eliminate the effects and uncertanties caused by the nuisance parameters in both two types of observational data sets (angular diameter distance $D_A$ and luminosity distance $D_L$). In this paper we further analyse the most updated QSO and SNe Ia data sets. Specially, we present a new approach that harnesses the ratio $\eta(z_i)/\eta(z_j)$ as cosmic observations, based on the luminosity distance inferred from the latest observations of type Ia supernova  and angular diameter distances obtained from the measurements of ultra-compact structure in radio quasars. All of the quantities used in the CDDR test come directly from observations, i.e., the absolute magnitudes of SN Ia and the linear size of the compact structure in radio quasars need not to be calibrated. In this way, the effects and uncertainties caused by the nuisance parameters are eliminated completely. More interestingly, our methodology will also benefit from the consistent redshift coverage of both samples that can reach a high redshift range of $z\sim2.3$.

With the increase both in the depth and quality of cosmological measurements, new techniques and methods for CDDR tests have also been developed, focusing on different machine learning algorithms \cite{2021PhRvD.103j3513A,2020JCAP...12..019H,2021MNRAS.504.3938M,2014PhRvD..89d3007M}. In this paper, we will use the Artificial Neural Network (ANN) algorithm to reconstruct the possible evolution of CDDR with redshifts. Note that such data-driven approach has no assumptions about the observational data, suggesting its advantage of being completely model-independent. Summarizing, we will propose an improved approach to test CDDR directly, which not only effectively avoids the influence of nuisance parameter on CDDR, but also achieves more stringent constraints on CDDR in the case of small data samples. The outline of this paper is given as follow: in Section II we briefly introduce the observations of ultra-compact structure of radio quasars acting standard rulers and SN Ia acting as standard candles. The improved methodology of testing CDDR and the corresponding results are presented in Section III.

\section{Observational data}

\subsection{Angular diameter distances from radio quasars}

We consider extracting angular diameter distance from angular redshift relation of compact structure of radio quasar. As the most distant and brightest objects in the Universe, quasars exhibit great potential in studying early cosmology beyond the limitation of supernovae. Unfortunately quasars exhibit large dispersion in luminosities at all wavelengths, which makes them unusable as standard probes for measuring cosmological distances. In the past decades, great effort have been made to make use of quasars as standard candles or standard rulers in modern cosmology, such as the Baldwin effect \citep{1977ApJ...214..679B}, the Broad Line Region radius-luminosity relation \citep{2011ApJ...740L..49W}, the properties of highly accreting quasars \citep{2013PhRvL.110h1301W}, and the non-linear relation between the ultraviolet and X-ray fluxes of the quasar to construct the Hubble diagram \cite{2019NatAs...3..272R,2015ApJ...815...33R,2017AN....338..329R,2020ApJ...899...71L,2020MNRAS.496..708L}. According to the unified model of active galactic nuclei and quasars, ultra-compact radio sources are identified as cases in which the jets are moving relativistically and are close to the line of sight. At any given frequency, the core is believed to be located in the region of the jet corresponding to unit optical depth with synchrotron self-absorption being the dominating process. In the original work of \citep{1993Natur.361..134K}, an interesting possibility was discussed that compact radio sources (especially quasars) constitute another potential class of standard rulers that could be observed by very long baseline interferometry (VLBI). The VLBI with high precision can not only accurately locate the radio source, but also measure the tension angle of the compact radio source at the magnitude of mas. Based on the subsequent works of \cite{1994ApJ...425..442G,2001CQGra..18.1159V}, the linear size of the compact structures in radio sources are related to the intrinsic luminosity $L$ and the redshift $z$ of the background source
\begin{equation}
l_m= lL^{\beta}(1+z)^n \label{l},
\end{equation}
where $l$ represents the linear size scaling factor which describes the apparent distribution of radio brightness within the core, $\beta$ and $n$ denotes the possible dependence of the intrinsic size on the luminosity and the redshift, respectively. However, the application of radio sources in cosmology still suffered from the high dispersion in the observed relations or the limitation of a poor statistics. With the gradually refined selection technique and observations, a key step forward was made in the work of  \cite{1994ApJ...425..442G}, which showed that the linear size dispersion in radio source with a flat spectral index $(-0.38<\alpha<0.18)$ is greatly reduced. Based on a sample of 2.29 GHz VLBI survey with 613 milliarcsecond compact radio sources, \cite{2017JCAP...02..012C,2017A&A...606A..15C} selected 120 intermediate-luminosity $(10^{27}$W/Hz$<L<10^{28}$W/Hz) quasars (ILQSOs) with reliable measurements on the angular size of the compact structure. The final results demonstrated that ILQSOs are almost independent from redshift and luminosity $(|n|\simeq 10^{-3}, \beta\simeq 10^{-4})$, which means they meet the requirements expected from standard rulers. However, the crucial question is what is the intrinsic metric linear size of the quasar source? The previous analysis roughly estimated that the $l_m$ parameter is robustly of the scale of $\sim 11$ pc \cite{2017JCAP...02..012C}. For the sake of the following description, we take a prior value $l_m=11.03\pm0.25$ pc determined by in a cosmological-model-independent method 
\cite{2017A&A...606A..15C}. However, in our work the value of $l_m$ does not affect the CDDR test, and we will later propose an improved CDDR test to eliminate the bias and additional systematic errors associated with the $l_m$ value of calibration.

\begin{figure}
\centering
{\includegraphics[width=8cm,height=6cm]{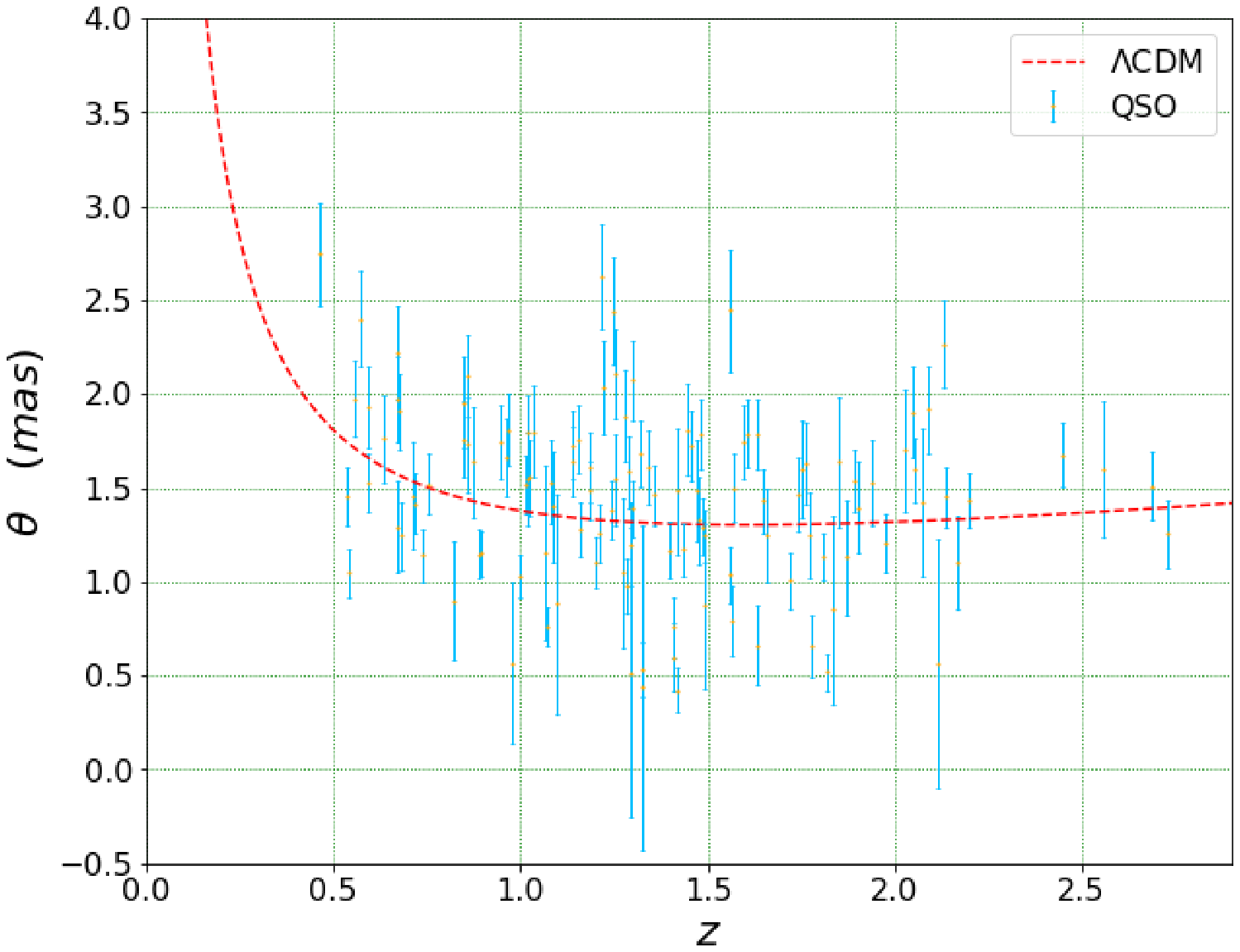}  \includegraphics[width=8cm,height=6cm]{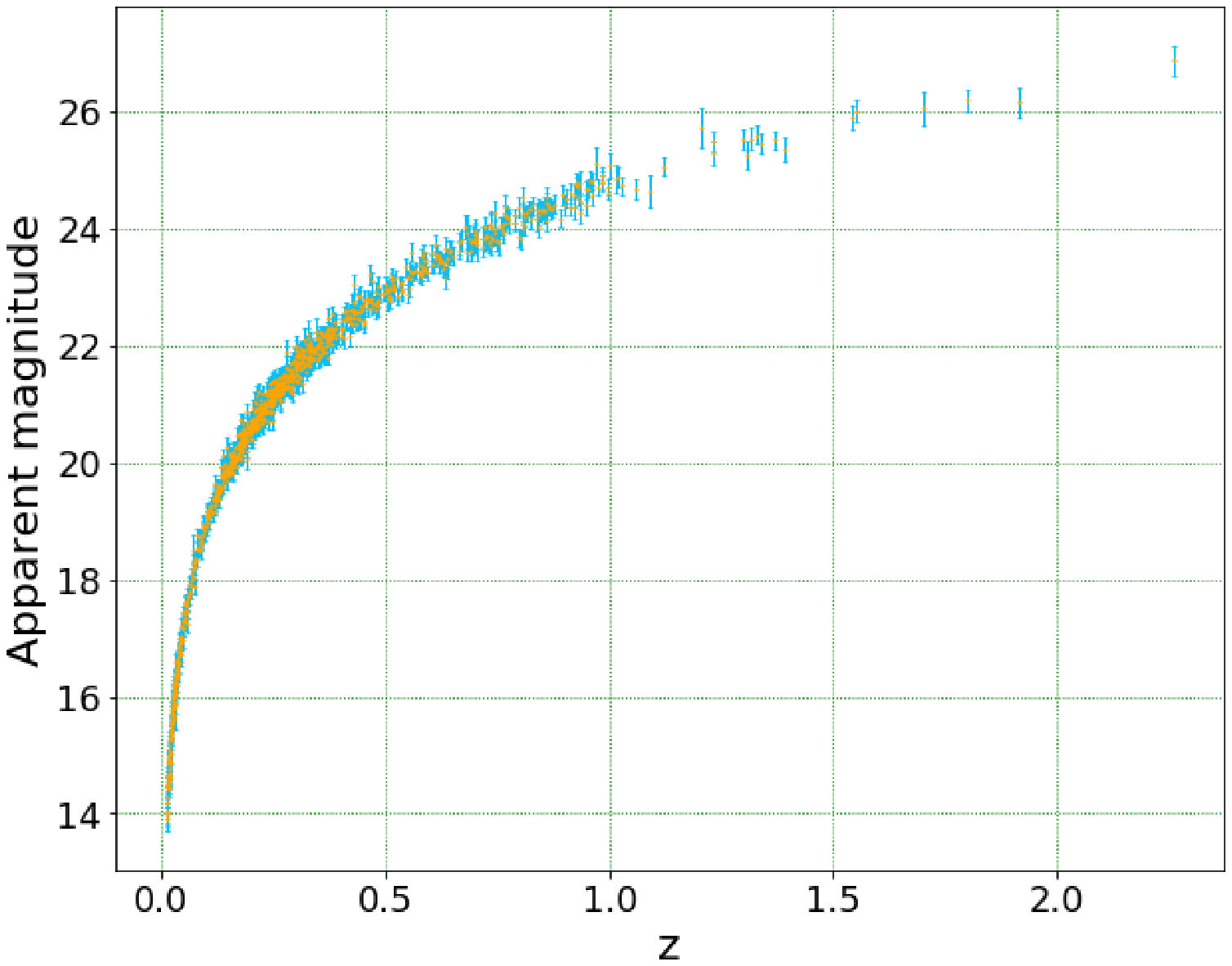}}
\caption{The scatter plot of the observed angular sizes of 120 radio quasars (left panel) and the apparent magnitudes of
1048 Pantheon SN Ia (right panel). The red dotted line denotes the angular sizes calculated from the fiducial $\Lambda$CDM model ($H_0=70.0$ km/s/Mpc, $\Omega_m=0.30$).}
\end{figure}
The angular size in compact structure and cosmic distance relation for cosmological inference was first proposed in \cite{1993Natur.361..134K}
\begin{equation}
D_A(z)= \frac{l_m}{\theta(z)}, \label{theta}
\end{equation}
where $D_A$ is the angular diameter distance, $l_m$ is the intrinsic metric linear size of the source, and $\theta(z)$ is the observed angular size, which is defined by the modulus of visibility $\Gamma=S_c/S_t$ in the literature \cite{1994ApJ...425..442G}. The specific definition of angular size is $\theta(z)=2\sqrt{-\ln\Gamma \ln 2}/\pi B_{\theta}$, where $B_{\theta}$ is interferometer baseline measured in wavelengths, $S_c$ and $S_t$ are correlated flux density and total flux density, respectively \cite{2017A&A...606A..15C}. The sample of raido quasars used in this work is the one described in \cite{2017JCAP...02..012C} with the redshift range between $z=0.462$ and $z=2.73$. These compact radio sources come from a well-known 2.29 GHz VLBI survey \cite{1985AJ.....90.1599P} (hereafter called P85) with 1398 detected candidates and 917 selected sources. The P85 sample was updated with respect to redshift \cite{2004JCAP...11..007J}, which includes 613 compact radio sources that cover the redshift range up to
$0.0035\leq z\leq 3.787$ (http://nrl.northumbria.ac.uk/13109/). These 120 radio quasars have been carefully selected for cosmological studies and we refer to \cite{2017JCAP...02..012C} for a detailed description of the selection procedure used to turn them into standard rulers and for an explanation of the calibration method used to include them in the extensive cosmological analysis \citep{2017arXiv170808867L,2017EPJC...77..891M,2017EPJC...77..502Q,2020ApJ...892..103Z}. The scatter diagram of the observed angular sizes for 120 radio quasars is shown in Fig.~1.

\subsection{Luminosity distances from Type Ia Supernova}

In order to carry out the test of CDDR, we need to find another cosmological probe that can directly provide luminosity distances and satisfy the following criteria, i.e., the probe should be able to cover roughly the redshift range of the compact radio quasars. In this work, we seek for SN Ia as a reasonably empirically well-understood cosmological probe, the usefulness of which to modern cosmology is well known in revealing the accelerated expansion of the Universe and placing constraints on cosmological parameters to break parameter degeneracies. With the rapid growth in the sample size of SN Ia distance measurements, the analysis and mitigation of systematic uncertainties of Type Ia Supernova has been considerably improved. However, the application of SN Ia for cosmology involves so-called "nuisance" parameters, which need to be optimized along with the unknown variables in cosmological models and could potentially affect reliable constraints on cosmological model parameters.

Fortunately, the recent SN Ia sample called Pantheon has been released by the Pan-STARRS1 (PS1) Medium Deep Survey, which contains 1048 SN Ia measurements spanning the redshift range $0.01<z<2.3$ \citep{2018ApJ...859..101S}. Here, we only summarise the crucial points required by the present work. Benefit from richness and depth of the sample,the Pantheon catalogue combines the subset of 279 PS1 SN Ia \cite{2014ApJ...795...44R,2014ApJ...795...45S} and useful distance estimations of SN Ia from SDS, SNLS, various low redshift and HST samples \citep{2018ApJ...859..101S}. More importantly, compared with the  previous SN Ia data sets \citep{2014A&A...568A..22B}, the Pantheon sample applies a new approach called BEAMS with Bias Corrections (BBC) \citep{2017ApJ...836...56K}, in which the apparent magnitude is replaced with the corrected apparent magnitude $m_{B,corr}=m_B+\alpha^{\star}\cdot X_1-\beta\cdot \mathcal{C}$ for all the SN Ia \cite{2018ApJ...859..101S}. Here, $m_B$ is the observed peak magnitude in rest-frame B band, while
$X_1$ and $\mathcal{C}$ are the color and light-curve shape parameters. The two nuisance parameters $\alpha^{\star}$ and $\beta$ should be fitted simultaneously with the cosmological parameters. It should be noted that the stretch luminosity parameter $\alpha^{\star}$ and the color-luminosity parameter $\beta$ are set to zero for the Pantheon sample. Therefore, the observed distance modulus of SN Ia provides the luminosity distance as 
\begin{equation}
D_{L,SN}(z)=10^{(m_{B,corr}(z)-M_B)/5-5} (Mpc),
\end{equation}
where $M_B$ is the absolute magnitude in B band. For the uncertainty of the luminosity distance in Pantheon data set, the contribution from photometric error, distance bias correction, and the peculiar velocity are included in this analysis \cite{2018ApJ...859..101S}. The apparent B-band magnitude for 1048 Pantheon SN Ia of is also illustrated in Fig.~1.

\begin{figure*}
\centering
\includegraphics[width=9cm,height=7cm]{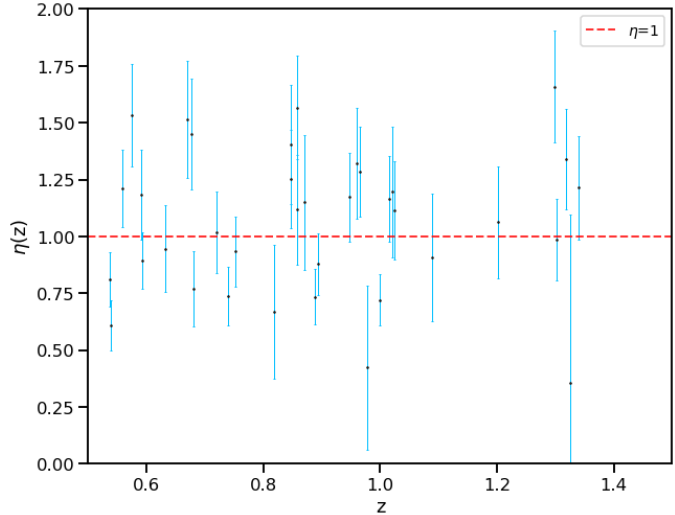}
\caption{The CDDR parameter $\eta(z)$ from the observations of radio quasars and SN Ia. }
\end{figure*}

\section{Methodology and results}

From the theoretical point of view, in order to directly test the DDR from observations, the following parameterized form is commonly used
\begin{equation}\label{eq:DDR}
\eta(z)=\frac{D_L(z)}{D_A(z)(1+z)^2},
\end{equation}
the likelihood of which is expected to peak at one in order to satisfy the CDDR. By combining Eqs.~(2) and (3) to Eq.~(4), one can rewrite the above expression as
\begin{equation}
\eta(z)=\frac{\theta(z)10^{(m_{B,corr}(z)-M_B)/5-5}}{l_m(1+z)^2}.
\end{equation}
The difficulty of testing CDDR lies in the fact that the angular diameter distance from an radio quasar should be observed at the same redshift as
SN Ia. In the previous work for example in \cite{2011RAA....11.1199C}, it was pointed out that the CDDR test could be significantly affected by the particular choice of the selection criteria for a given pair of data sets. Following the redshift selection criterion widely used in the literature (within the redshift range of $0.01<z<2.3$) \cite{2011ApJ...729L..14L,2016ApJ...822...74L,2021EPJC...81..903L}, the redshifts of SN Ia sample are carefully chosen to coincide with the associated quasar sample demanding that the difference in redshift is smaller than 0.005. By performing such selection criterion that could hopefully ease the systematic errors brought by redshift inconsistency, only 37 pairs of data sets are obtained from the Pantheon and ILQSO sample. Combining these quasar data together with the Pantheon SN Ia sample, we obtain the CDDR parameter $\eta(z)$ shown in Fig.~2. More specifically, the total uncertainties of $\eta(z)$ are calculated from the standard uncertainty propagation formula, based on the uncorrelated uncertainties of observables including the observed angular size errors $\sigma_{\theta}$, corrected apparent magnitude errors $\sigma_{m_{B, corr}}$, as well as additional systematic errors introduced from the calibrations of absolute magnitude ($M_B$) of SN Ia and linear size ($l_m$) of radio quasars. To better illustrate the statistical significance of our results, we first use the weighted mean statistics \citep{1993ComPh...7..415B} to evaluate  
\begin{equation}
\eta=\frac{\Sigma_i\big(\eta_i/\sigma_{\eta_i}^2\big)}{\Sigma_i\big(1/\sigma_{\eta_i}^2\big)},\,\,\,\,\,\,\,\,\,
\sigma^2_{\eta}=\frac{1}{\Sigma_i\big(1/\sigma_{\eta_i}^2\big)},
\end{equation}
where $\eta$ stands for the weighted mean and $\sigma_{\eta}$ is its corresponding uncertainty of CDDR parameter. Such statistical method has been widely applied in meta-analysis to integrate the results of independent measurements \citep{2019NatSR...911608C}. Our assessments for weighted mean and corresponding uncertainty are $Mean(\eta(z))=0.991(\pm0.147)$, which is in perfect agreement with the results of previous works \citep{2016ApJ...833..240L,2017ApJ...838..160W,2021PhRvD.103j3513A,2020JCAP...12..019H,2021MNRAS.504.3938M}, indicates that there is no evidence for the CDDR violation. Given the possible invalidity of Gaussian distribution of the errors, we also use a robust median statistics \citep{2012msma.book.....F} to evaluate the measurements of $\eta(z)$. Moreover, if there are extreme values and outliers in the sequence, it is better to use the median as the representative value. When making a total number of $N$ measurements, one might naturally expect that there is a 50\% chance that each measurement is higher/lower than the true median. Therefore, the probability that $n$-th observation is higher than the median follows the binomial distribution: $P=2^{-N}N!/[n!(N-n)!]$ \citep{2001ApJ...549....1G}. Similarly, we can define the  68.3$\%$ confidence interval with median statistics. In the framework of such non-parametric approach, the resulting constraint on the CDDR parameter becomes $Med(\eta(z))=1.117(\pm0.328)$ with the median value and the absolute deviation. Therefore, the conclusion of CDDR validity ($\eta(z)=1$) seems robust within $1\sigma$ confidence interval.

\begin{figure*}
\centering
{\includegraphics[width=8cm,height=6cm]{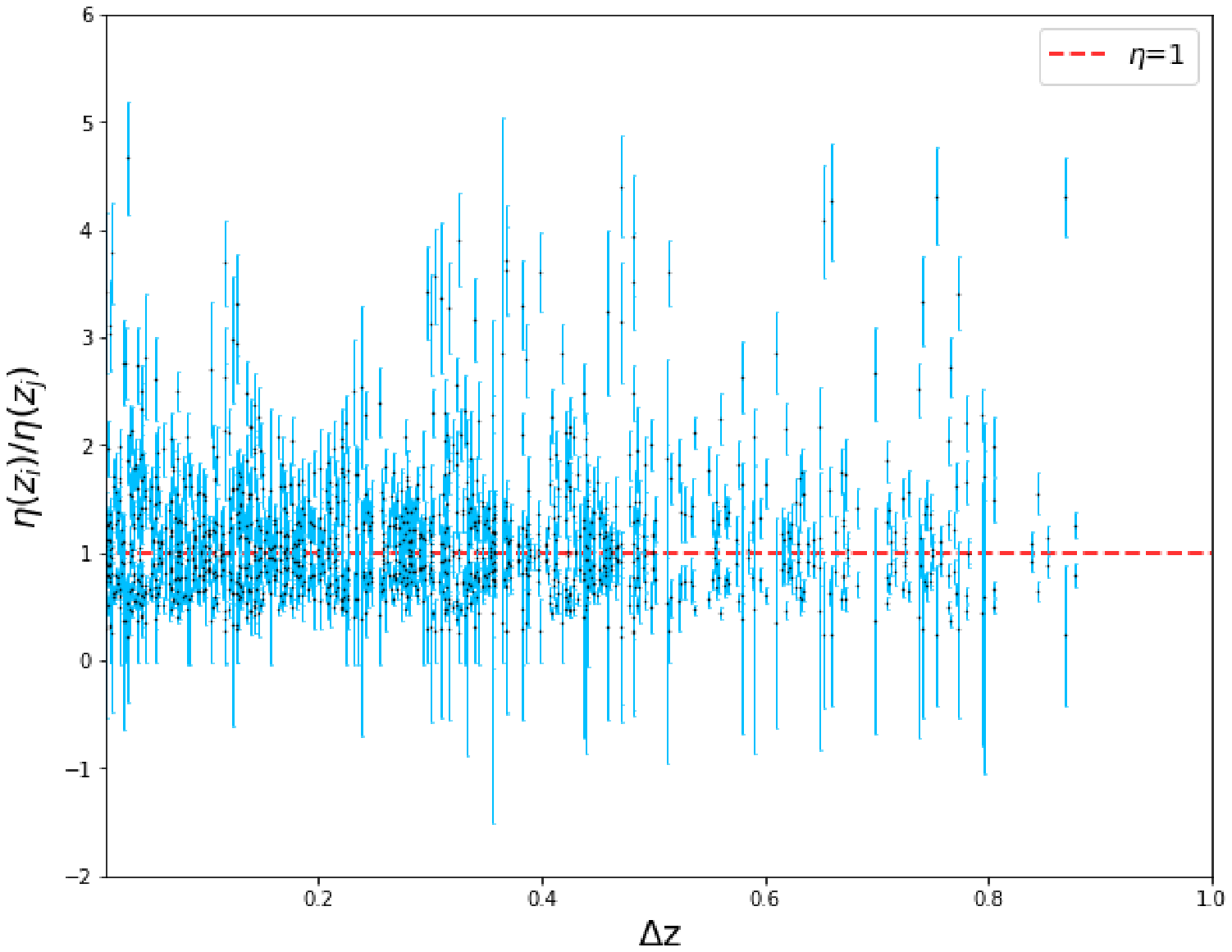}
\includegraphics[width=8cm,height=6cm]{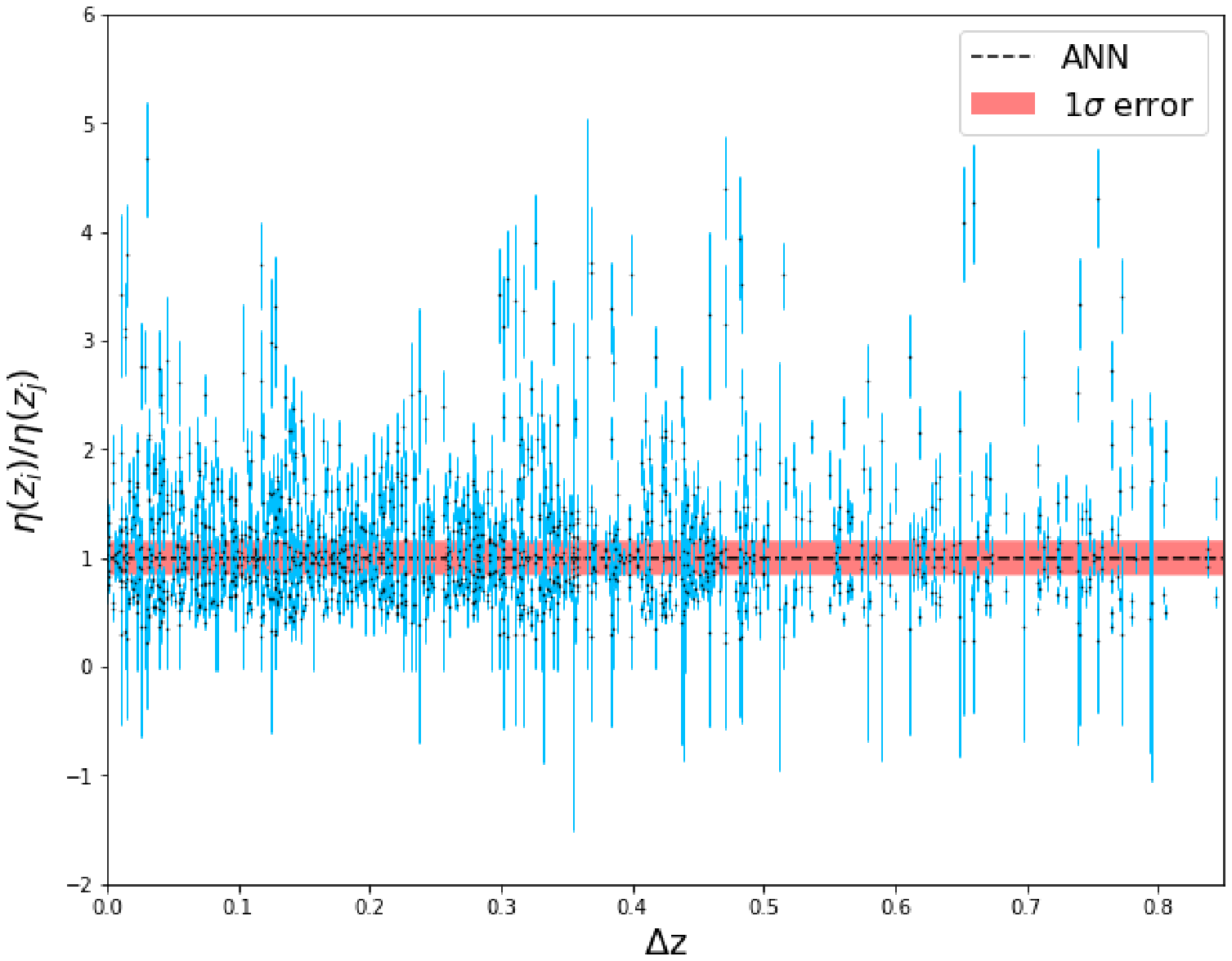}}
\caption{The $\eta(z_i)/\eta(z_j)$ two-point diagnostics calculated on the observations of radio quasars and SN Ia (left panel). The reconstructed $\eta(z_i)/\eta(z_j)$  two-point diagnostics with ANN machine learning algorithm (right panel).}
\end{figure*}

\begin{figure}
\centering
{\includegraphics[width=7.5cm,height=6.2cm]{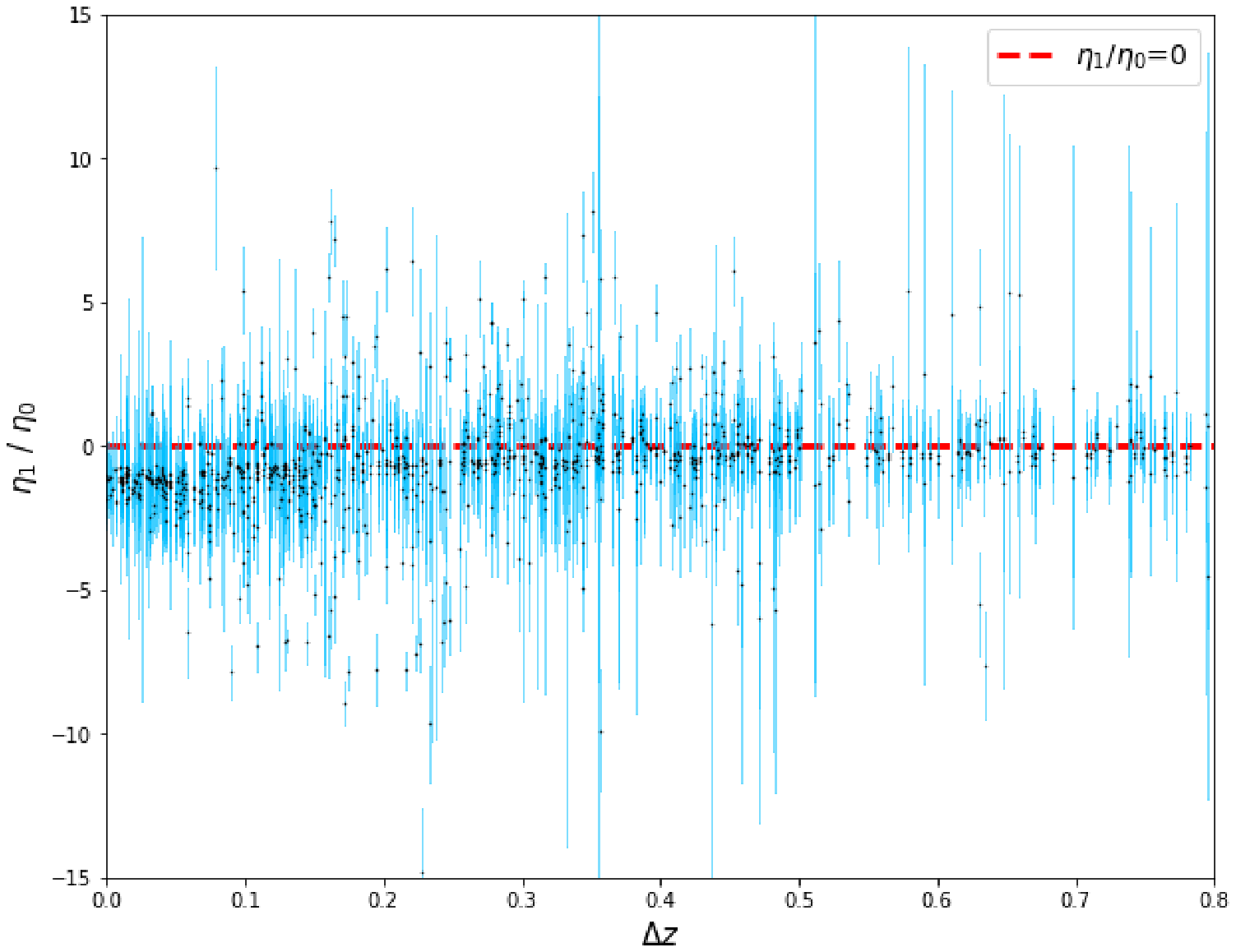}
\includegraphics[width=7.5cm,height=6.4cm]{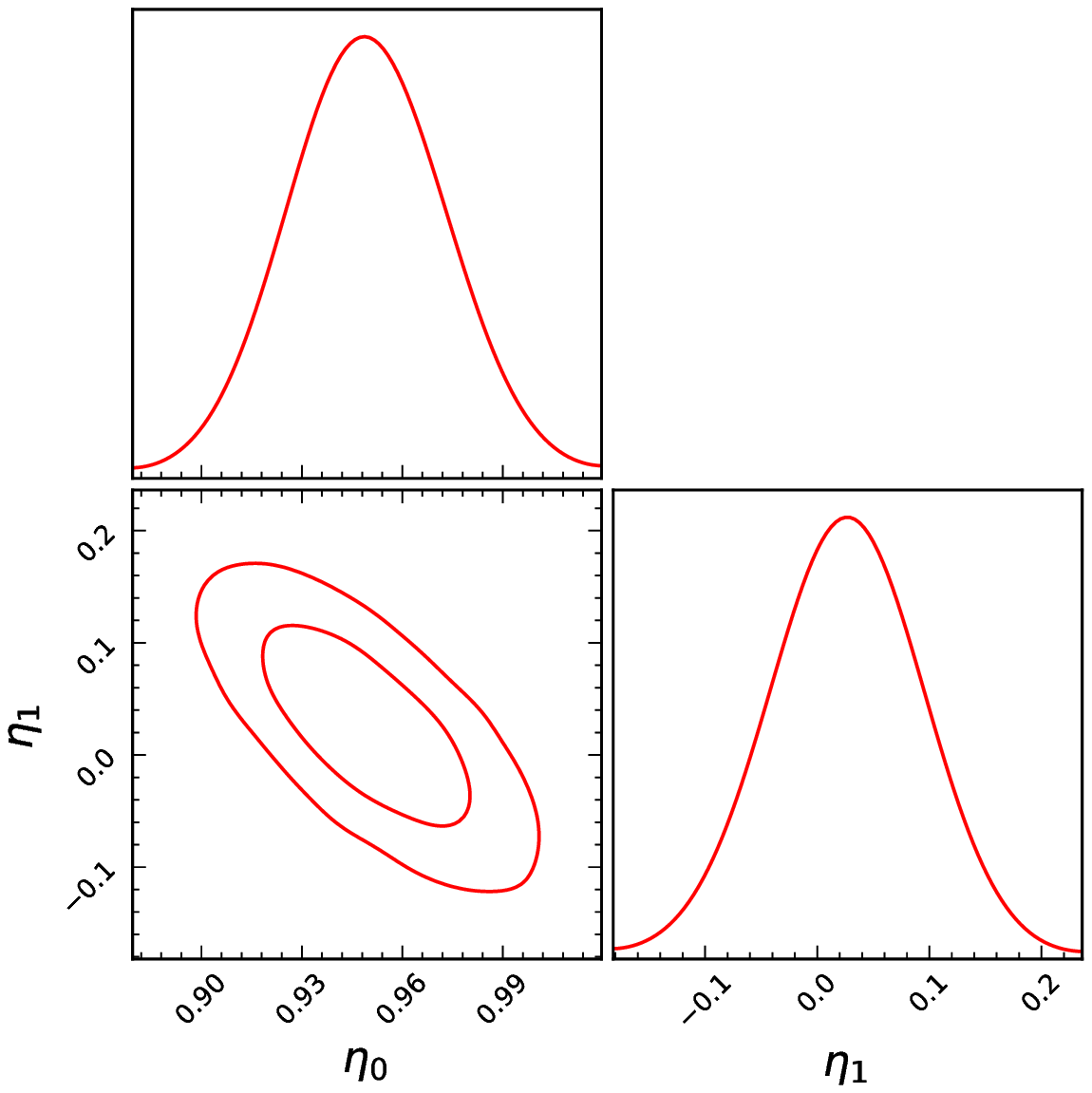}}
\caption{The scatter plot of the CDDR parameter $\eta_1/\eta_0$ (left panel) and constraints on the CDDR parameters $(\eta_0, \eta_1)$  (right panel), in the framework of $\eta(z_i)/\eta(z_j)$ two-point diagnostics.}
\end{figure}

Due to the ambiguous interpretation of the compact structure size in radio quasars and the absolute B-band magnitude of SN Ia whose value is determined by the host stellar mass, the linear size parameter $l_m$ and the absolute magnitude $M_B$ are hard to determine precisely. In fact, the uncertainty of CDDR measurements shown in Fig. 2 is dominated by the calibration of two nuisance parameters. In order to eliminate the influence of these two nuisance parameters, we propose an improved approach by introducing the ratio of CDDR parameter
\begin{equation}
\eta(z_i)/\eta(z_j)=\frac{\theta(z_i)(1+z_j)^2}{\theta(z_j)(1+z_i)^2}
10^{\Delta m_{B,corr}/5},
\end{equation}
where $\Delta m_{B,corr}=m_{B,corr}(z_i)-m_{B,corr}(z_j)$ is the difference of corrected apparent magnitude between arbitrary two SN Ia data points. If one defines the ratio $\eta(z_i)/\eta(z_j)$, where $i, j$ denote
the order numbers of the radio quasars and SN Ia, then such quantity does not depend on the nuisance parameters and it does not introduce any uncertainty to the results. Note that 
if we have observational data at $n$ different redshifts, then we can get $n(n-1)/2$ data pairs. The uncertainty of $\eta_{ij}=\eta(z_i)/\eta(z_j)$ is calculated using the standard error propagation formula, which is related to the uncorrelated uncertainties of the observed angular size $\sigma_{\theta}$ and corrected apparent magnitude $\sigma_{m_{B, corr}}$. More importantly, our approach successfully eliminate the nuisance parameters $M_B$ and $l_m$, which brings benefits in alleviating the systematics caused by precise determination of these parameters. These are the apparent merits of our methodology. Our approach was inspired by the two-point diagnostic approach, which has been extensively applied to quantify the difference between the cosmological constant ($\Lambda$CDM) and other dark energy models (including
evolving dark energy) \cite{2012PhRvD..86j3527S,2016ApJ...825...17Z,2018arXiv180309106Z}.

In order to gain insight concerning the two-point diagnostics calculated for every combination of pairs taken from the full QSO+SN Ia data.  We  display these diagnostics together with their uncertainties as a function of redshift difference $\Delta z=|z_i-z_j|$ in the left panel of Fig.~3. Benefit from the improved methodology, the QSO/SN Ia pairs satisfying irrespective of the redshift selection criteria have a massive growth. One can see that there are some interesting features regarding the uncertainties of the two-point diagnostics, i.e., they are apparently non-Gaussian. In order to test further the validity and efficiency of our method, we use two approaches to produce a summary statistics of two-point diagnostics calculated on the data sets. The first is to use the weighted mean statistical method. In order to ensure that each data point is uncorrelated, the weighted mean formula for the $\eta(z_i)/\eta(z_j)$ diagnostic should be rewritten as \cite{2016ApJ...825...17Z}
\begin{equation}
\eta_{ij}=\frac{\Sigma_{i=1}^{n-1}\Sigma_{j=i+1}^{n}\big(\eta_{ij}/\sigma_{\eta_{ij}}^2\big)}
{\Sigma_{i=1}^{n-1}\Sigma_{j=i+1}^{n}\big(1/\sigma_{\eta_{ij}}^2\big)},\,\,\,\,\,\,\,\,\,
\sigma^2_{\eta_{ij}}=\frac{1}{\Sigma_{i=1}^{n-1}\Sigma_{j=i+1}^{n}\big(1/\sigma_{\eta_{ij}}^2\big)}.
\end{equation}
The weighted mean value and corresponding uncertainty is $Mean(\eta(z_i)/\eta(z_j))=0.968\pm0.031$, which suggests that the weighted mean of this diagnostic is compatible with CDDR within the observational uncertainty. Actually, benefit from the absence of nuisance parameters involved in other currently available methods, our methodology produces more stringent constraints on CDDR (with the precision of $10^{-2}$) at the current observational data level. The second approach is the median statistics method, which is an appropriate measure in light of the non-Gaussian error distribution. The validity of CDDR at $z\sim 2.3$, with the 68\% confidence intervals of the median $Med(\eta(z_i)/\eta(z_j))=0.998(\pm0.436)$, seems much more justified than the previous one drawn from the weighted mean. Therefore, the results of $\eta(z_i)/\eta(z_j)$ showed in this paper demonstrate no evidence for the deviation from CDDR irrespective of the statistical method used. This is one of the unambiguous conclusions in our work. However, one should also be aware of the disadvantage of the above method, i.e., the ratio of CDDR parameter $\eta(z_i)/\eta(z_j)$ should be constant and exactly equal to one if the CDDR is the true one. However, the CDDR can be violated even if the ratio is exactly equal to one. In order to fully explore the consequences of our proposed $\eta(z_i)/\eta(z_j)$ diagnostics, we adopt an explicit parameterization $\eta(z)=\eta_0+\eta_1z$ to better illustrate what our results imply for the redshift-evolution of CDDR parameter. Thus, the ratio of CDDR parameter can be rewritten as
\begin{equation}
\frac{\eta_1}{\eta_0}=(\frac{\Delta z}{1-\frac{\theta(z_i)(1+z_j)^2}{\theta(z_j)(1+z_i)^2}
10^{\Delta m_{B,corr}/5}}-z_j)^{-1},
\end{equation}
which should be equal to zero if there is no redshift evolution of CDDR. The measurements of these diagnostics as a function of redshift difference $\Delta z$ are shown in Fig.~4. Furthermore, we also use a Python Markov Chain Monte Carlo (MCMC) module \citep{2013PASP..125..306F} to obtain fits on the two CDDR parameters, by minimizing the $\chi^2$ objective function 
\begin{equation}
\chi^2=\frac{2}{n(n-1)}\sum_{i=1}^{n-1}\sum_{j=i+1}^{n}\frac{(\eta_{ij}^{th}-\eta_{ij}^{obs})}{\sigma^2_{\eta_{ij}}}.
\end{equation}
In Fig.~4 we also plot the one-dimensional marginalized distributions and two-dimensional constraint contours for the CDDR parameters, with the best-fit values of $\eta_0=0.952^{+0.019}_{-0.019}$ and $\eta_1=0.023^{+0.053}_{-0.054}$, respectively. 
It is worth to comment that on the one hand, our methodology produces a possible deviation from the expected value of CDDR parameter ($\eta_0=1$) up to $z\sim 2.3$. However, our results are still marginally consistent with the CDDR validity within $2\sigma$ C.L., which is in full agreement with other recent tests involving cosmological data. A summary of the current constraints on the $\eta_0$ from different cosmological observables can be found in Ref.~\cite{2022EPJC...82..115H}. On the other hand, the CDDR remains redshift independent ($\eta_1=0$) within $1\sigma$ C.L., supporting the persisting claims that the Etherington reciprocity theorem could still be the best description of our universe.

\begin{figure*}
\centering
{\includegraphics[width=8cm,height=6cm]{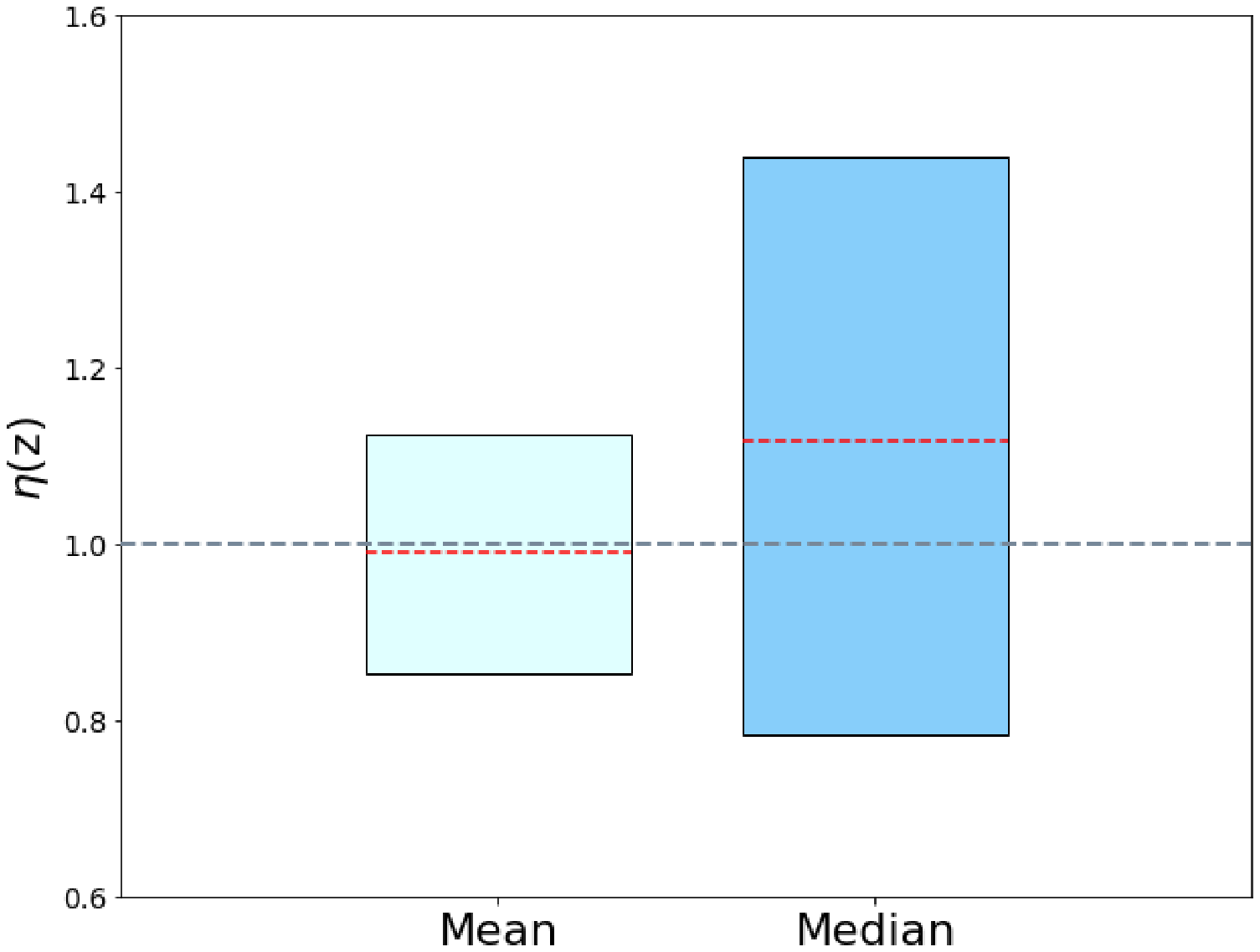}\\
\includegraphics[width=8cm,height=6cm]{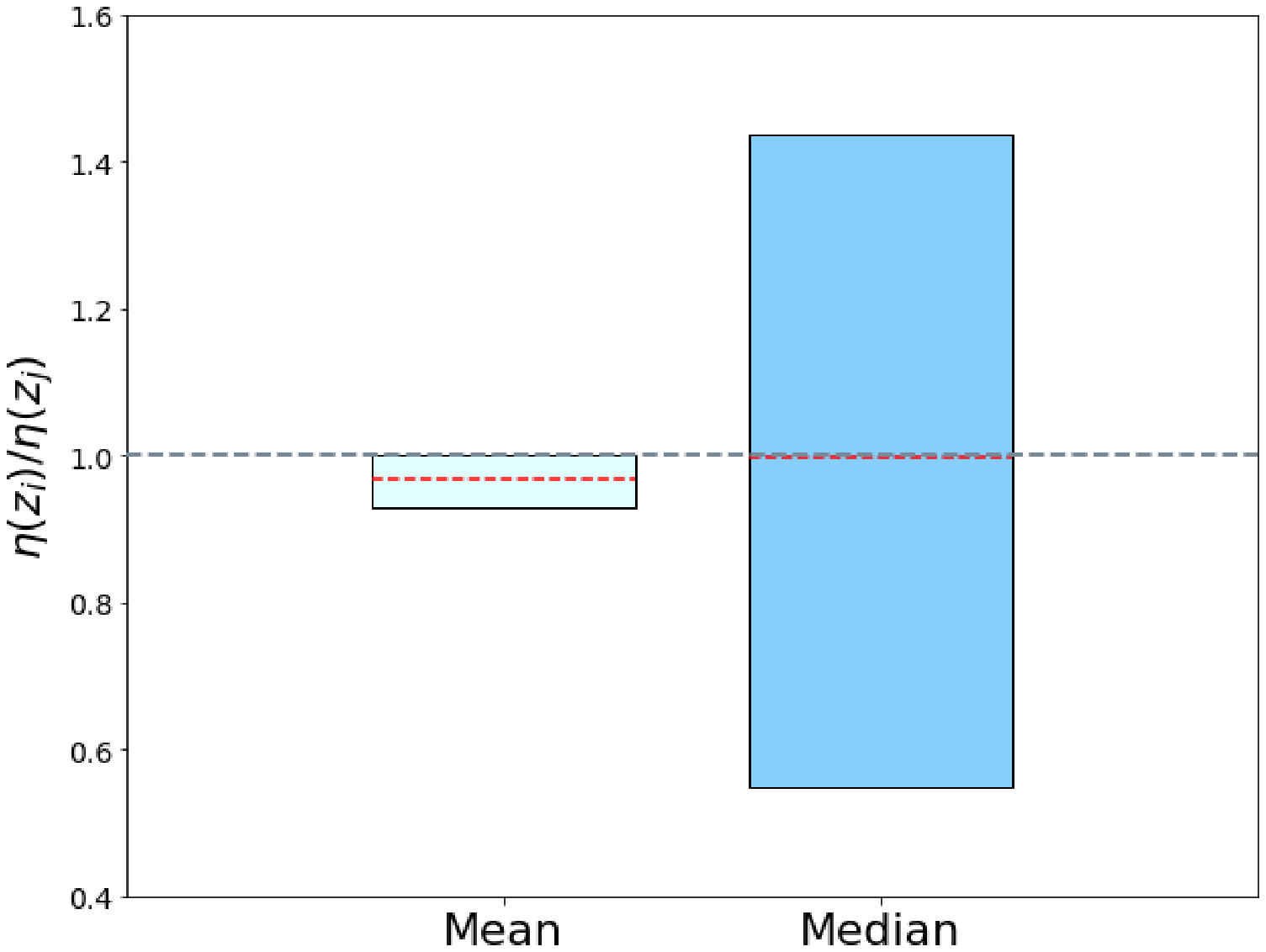}
\includegraphics[width=8cm,height=6cm]{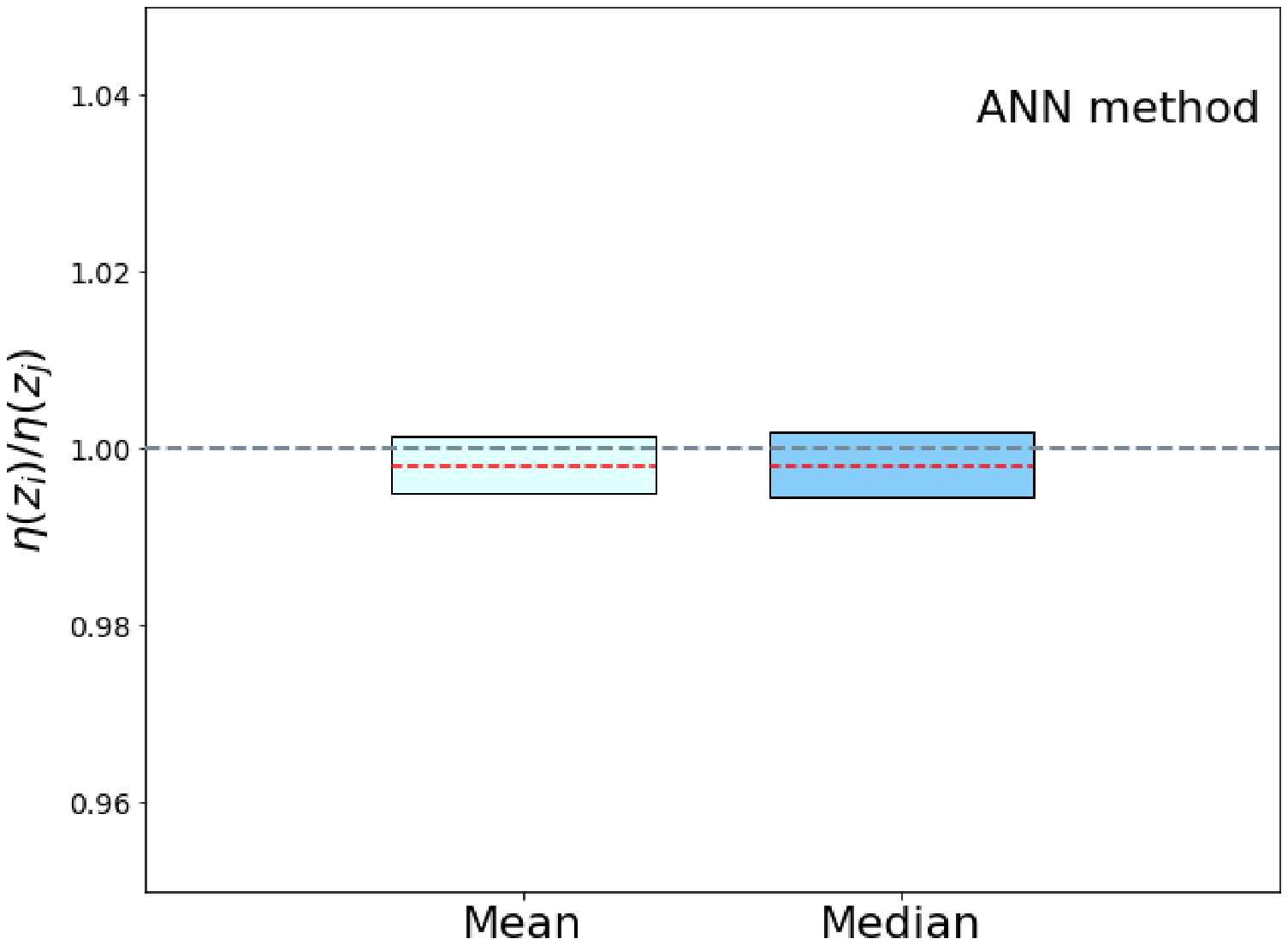}}
\caption{The CDDR parameters $\eta(z)$ and $\eta(z_i)/\eta(z_j)$ calculated from the two statistical methods as weighted mean and the median statistics. Bands display the 68.3\% confidence regions.}
\end{figure*}

There are many ways the above findings could be improved. For instance, it is still interesting to see whether those conclusions may be changed with machine learning algorithms, which have shown their excellent potential in addressing cosmological issues and constraining cosmological parameters \cite{2018PhRvD..98l3518F,2019PhRvD.100f3514F,2019MNRAS.490.1843R,2020ApJS..249...25W}. More importantly, as a completely data driven approach, the Artificial Neural Network (ANN) method does not assume random variables that satisfy the Gaussian distribution. The main purpose of an ANN (which consists of an input layer, one or more hidden layers and an output layer) is to construct an approximate function $f_{\mathbf{W},\mathbf{b}}(\mathbf{x})$ (in which $\mathbf{W}$ and $\mathbf{b}$ are linear weights matrix and the offset vector) that correlates the input vector $\mathbf{x}$ with the output vector $\mathbf{y}$ \cite{2015arXiv151107289C}. According to the difference between the predicted value $f_{\mathbf{W},\mathbf{b}}(\mathbf{x})$ of the current network and the target value $\mathbf{y}$, the weight matrix of each layer needs to be constantly updated for minimize the difference, which is defined by a loss function $\mathcal{L}$ \cite{2014arXiv1412.6980K}. An issue that needs clarification
is the achievable $1\sigma$ confidence region for the reconstructed function, which depends on both the actual errors and the cost function. Following the detailed discussion in \cite{2021MNRAS.501.5714W}, a complete artificial neural network has the following parts: firstly, the weight is randomly initialized in the neural network; Secondly, the output value is compared with the expected output value, and the cost function is used to calculate the error; Thirdly, the error is propagated back to the neural network and the weight is set according to this information; Fourthly, repeat steps two to four for each input value in the training set; Finally, when the entire training set is sent to the neural network, the entire training is complete. The recent analysis has demonstrated the effectiveness of ANN acting as ``universal approximator" to produce representative uncertainties of the observations, especially in high-precision test of CDDR in both electromagnetic and gravitational wave domain \cite{2021EPJC...81..903L}. In particular, \textit{Euclid} collaboration improved the precision of CDDR test by approximately a factor of six, based on machine learning reconstruction using genetic algorithms \cite{2020A&A...644A..80M}.

Using the publicly released code called Reconstruct Functions with ANN \footnote{ReFANN; https://github.com/Guo-Jian-Wang/refann.}, we perform the reconstruction of the parameter $\eta(z_i)/\eta(z_j)$ based on the current  $\eta(z_i)/\eta(z_j)$ two-point diagnostics. The reconstructed functions with corresponding 1$\sigma$ uncertainties, which can be considered as the average level of observational error are given in right panel of Fig.~3. Working on the reconstructed 1000 $\eta(z_i)/\eta(z_j)$ points with ANN, we obtain $Mean(\eta(z_i)/\eta(z_j))=0.998(\pm0.003)$ and $Med(\eta(z_i)/\eta(z_j))=0.998(\pm0.004)$ in the framework of weighted mean and median statistics. Therefore, with ANN algorithm one could expect the parameter $\eta(z_i)/\eta(z_j)$ to be estimated at the precision of $10^{-3}$, which is more stringent than other results based on currently available observational data. In order to facilitate comparison between the inferred values of CDDR parameters obtained from two statistical approaches, we display the results in Fig.~5. As a final remark, possible violations of such fundamental relation (cosmic distance duality relation) might have profound implications for the understanding of fundamental physics and natural laws. Based on better uv-coverage in the future, we pin our hope on multi-frequency VLBI observations of more compact radio quasars with higher angular resolution, smaller statistical and systematic uncertainties. Meanwhile, considering the variety of different machine learning algorithms, we may also be optimistic in detecting possible deviation from the CDDR with much higher precision.

\section{Acknowledgments}
This work was supported by the National Natural Science Foundation of China under Grants Nos. 12203009, 12122504, 12021003, 11875025, 11633001, 11920101003, and  62202469; the Strategic Priority Research Program of the Chinese Academy of Sciences, Grant No. XDB23000000; Beijing Natural Science Foundation (Grant No. 4224091); the Interdiscipline Research Funds of Beijing Normal University; and the China Manned Space Project (Nos. CMS-CSST-2021-B01 and CMS-CSST-2021-A01).

\bibliographystyle{unsrt}
\bibliography{ddr_PLB}

\end{document}